\documentstyle[11pt,dunk2001_asp,twoside,psfig]{article}
\newcommand{\sm}{\, {\rm M}_{\odot}}

\def\edcomment#1{\iffalse\marginpar{\raggedright\sl#1\/}\else\relax\fi}
\reversemarginpar

\begin{document}
\title{On the phase-space structure of the Milky Way dark-matter halo}
 \author{Amina Helmi, Volker Springel \& Simon D.M. White}
\affil{Max Planck Institut  f\"ur Astrophysik, Karl
 Schwarzschildstr. 1, 85741 Garching bei M\"unchen, Germany}

\begin{abstract}
We analyse a high resolution simulation of the formation of a
cluster's dark-matter halo in a $\Lambda$CDM cosmology (Springel et
al. 2001).  The resolution achieved allows us to map the phase-space
structure in detail, and characterize its evolution and degree of
lumpiness.  Scaling down the cluster halo to a Milky-Way size
halo, we probe the substructure expected in the solar neighbourhood.
Here we specifically address the relevance of such substructure for
direct detection experiments aimed at determining the nature of
dark-matter.
\end{abstract}

\vspace*{-1cm}\section{Introduction}

Over the last twenty years a theory has emerged for the formation of
structure in the Universe (Peebles 1974; White \& Rees 1978). The
hierachical paradigm has allowed astronomers to make very definite
predictions for the properties of galaxies today and about their
evolution from high redshift. Direct comparisons to observations have
shown that this model is quite successful in reproducing both the
local and the distant Universe. However, it relies on a basic {\it
assumption}: that most of the matter in the Universe is dark, in the
form of some yet to be identified weakly interacting elementary
particles.

Thus the crucial test of this paradigm consists in the determination
of the nature of dark-matter through direct detection
experiments. Among the most promising candidates from the particle
physics perspective are axions and neutralinos. 
Axions can be detected
through their conversion to electric photons in the presence of a
strong magnetic field. The most important direct detection process of
neutralinos is elastic scattering on nuclei, and the idea is to
determine the count rate over recoil energy above a given (detector)
background level.  
The experimental situation has been improving
rapidly over the past years, as the relevant regions of parameter
space for the different dark-matter candidates are starting to be
probed (Bergstr\"om 2000).  In all experiments, the count rate is
strongly dependent on the velocity distribution of the incident
particles. In most cases, an isotropic Maxwellian distribution has
been assumed (e.g. Freese, Frieman \& Gould 1988), although there are
other examples in the recent literature, discussing multivariate
Gaussian distributions (e.g. Green 2000). Attempts at understanding
the effect of substructure on the velocity distribution of dark-matter
particles have also been made (e.g. Hogan 2001). However, most of this
work did not assume realistic distributions of matter in the Solar
neighbourhood, as we shall demonstrate in the next section.

The simulations we analyse here were generated by zooming in and
re-simulating with higher resolution a particular cluster and its
surroundings formed in a cosmological $\Lambda$CDM simulation, with
parameters $\Omega_0 = 0.3$, $\Omega_\Lambda = 0.7$, $h= 0.7$ and
$\sigma_8 = 0.9$. The cluster selected for resimulation is the second
most massive cluster in the parent simulation and has a virial mass of
$8.4 \times 10^{14} h^{-1} \,{\rm M}_{\odot}$ (see the left panel of
Figure~1). The highest resolution resimulation of the cluster region
has $6.6 \times 10^7$ particles. We scale the cluster to a ``Milky Way"
halo by requiring that the maximum circular velocity is 220 km/s,
which yields a scaling factor $\gamma= v_c^{cl}/v_c^{MW} \sim 9.18$.

\section{The phase-space structure in the Solar neighbourhood}
\subsection{Spatial distribution}

One of the critical issues in understanding the outcome of the various
dark-matter experiments consists in understanding the expected signal:
Is the distribution of particles in the vicinity of the Sun smooth
or is it dominated by a just a few streams or even bound lumps
(e.g. Moore et al.  2001)?  In Figure~1 we plot the positions of all
particles inside a cubic volume of 2 kpc on a side, located at the
position of the ``Sun". The spatial distribution of particles inside
this representative volume is extremely smooth. This is mostly due to
the fact that the material that ends up in the inner galaxy mostly
comes from a few very massive halos which have rapidly mixed.
\begin{figure}
\centerline{\hbox{
\psfig{figure=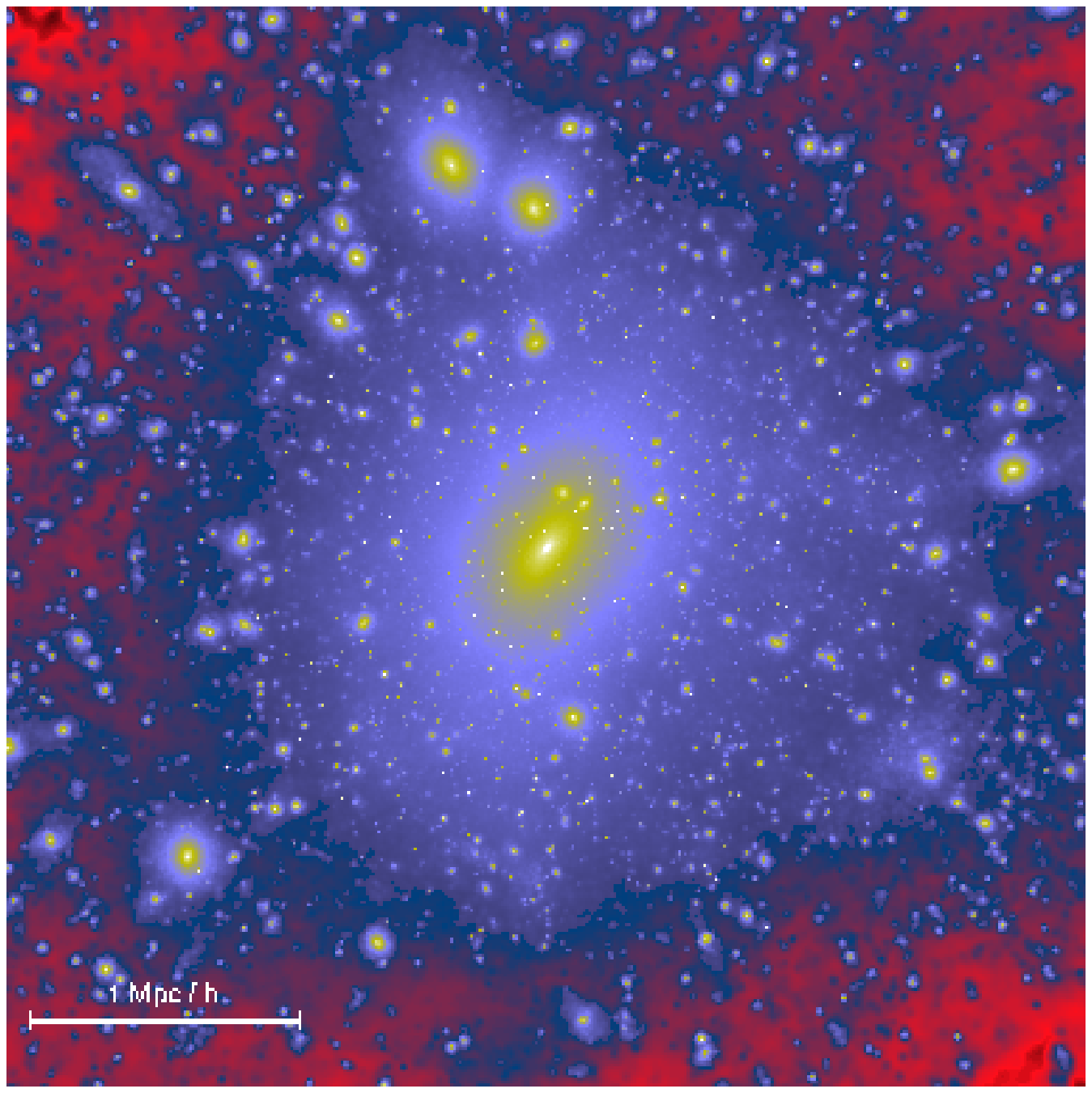,height=6cm}
\psfig{figure=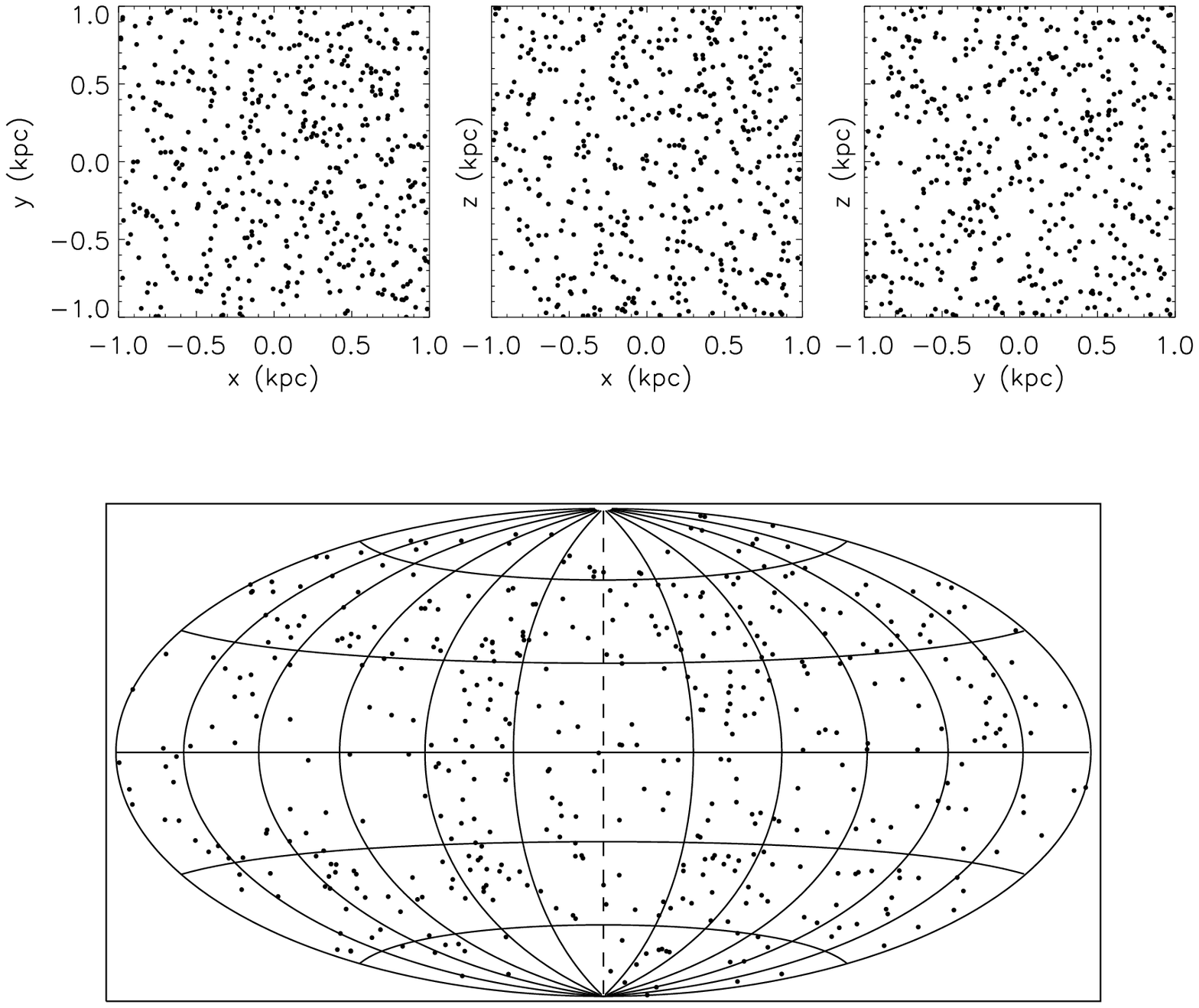,height=6cm}}}
\caption{The left panel shows the projected mass distribution of the
simulated halo out to the virial radius. The top panels on the
right show the spatial distribution of particles inside a 2
kpc on a side volume located at 8 kpc from the halo centre, i.e. this
volume is centered on the ``Sun".  The bottom panel shows their
distribution in the sky.}
\end{figure}

\subsection{Kinematics}

 In Figure~2 we show the velocities of particles located in a box of 4
kpc on a side in the vicinity of the Sun. Their velocity distribution
is relatively smooth, and appears to be quite consistent with a
Gaussian (see left panel Figure~3), at least to the ``naked eye".
However if we focus on the highest energy particles this seems no
longer to be the case, as shown by the particles highlighted in
grey. The 1\% fastest moving particles are strongly clumped,
and their velocity distribution is highly anisotropic.
\begin{figure*}
\centerline{\psfig{figure=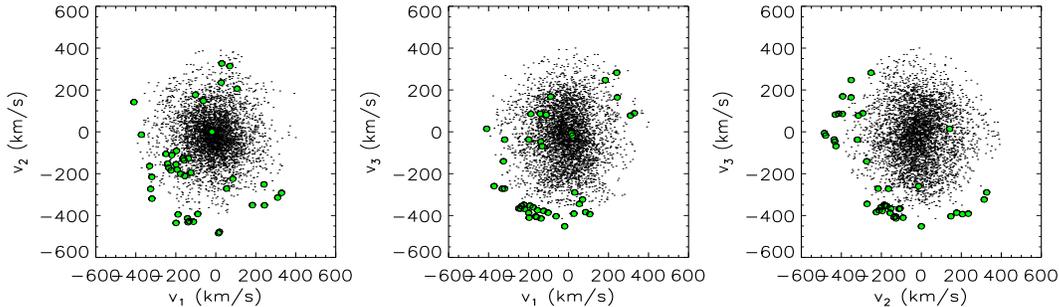,width=14cm,height=4cm}}
\caption{Velocities of particles located in a box of 4 kpc on a side
at the ``Solar" radius. There are 4362 particles in this volume, and
highlighted are the 1\% fastest. The velocity dispersions along the
principal axes are $(111.2, 120.1,141.4)$ km/s.  The lump with $v \sim
( -250,-200, -400)$ km/s corresponds to a halo of $1.94 \times 10^{10}
\sm$ identified at $z= 2.4$, and accreted at $z=1$, or $8.2$ Gyr ago.}
\end{figure*}

To quantify the substructure present in this volume we compute 
the correlation function $\xi$ in velocity space, defined as
$\xi = \frac{\langle DD \rangle }{\langle RR \rangle } - 1$, 
where $\langle DD \rangle$ is the number of pairs of particles in our
simulation with velocities in a given velocity range (or bin), and
$\langle RR \rangle$ is defined analogously for the same number of
random points, which we draw from a trivariate Gaussian distribution
determined from the data in the principal axes velocity frame, and is
the average over ten such realizations. The right panel in Figure~3
shows that there is a weak excess of particles with similar velocities
(i.e. below 100~km/s) with respect to what would be expected for a
random multivariate Gaussian distribution.  However, if we focus on the
fastest moving particles, the excess of particles is very noticeable,
particularly at small velocity differences, indicative of the presence
of streams, as clearly visible in Fig.~2.

\vspace*{-0.35cm}
\section{Conclusions}

A dark-matter halo formed in a $\Lambda$CDM cosmology is not a smooth
entity. Not only do dark-matter halos contain a large number of dark
satellites, they also have large amounts of substructure in the form
of streams.  By scaling a cluster halo down to a galaxy halo using the
ratio of their maximum circular velocities (equivalent to scaling
their mass ratio to the one-third power), we predict that the Galactic
dark-matter halo in the ``Solar neighbourhood" should be smoothly
distributed in space. This material is clumped in at least a few
hundred kinematically cold streams, as more detailed analyses show
(Helmi, White \& Springel 2001). These streams have their origin in
the different halos that merged to form the dark halo of the galaxy.

Our results indicate that direct detection experiments may quite
safely assume a multivariate Gaussian to represent the motions of
the dark-matter particles. However, experiments which are only
sensitive to the highest energy (i.e. fastest moving) dark-matter
particles will be subject to the presence of a few dominating streams,
and an excess of particles with similar energies should then be
expected in comparison with a smooth Gaussian distribution.  If direct
detection experiments are also sensitive to the direction of motion of
the incident dark-matter particles, the signal expected for the
fastest moving particles is also highly anisotropic, and could be
eventually be used, not just to determine the nature of the
dark-matter, but also to recover at least partially the merging
history of our galaxy.
\begin{figure}
\centerline{\hbox{
\psfig{figure=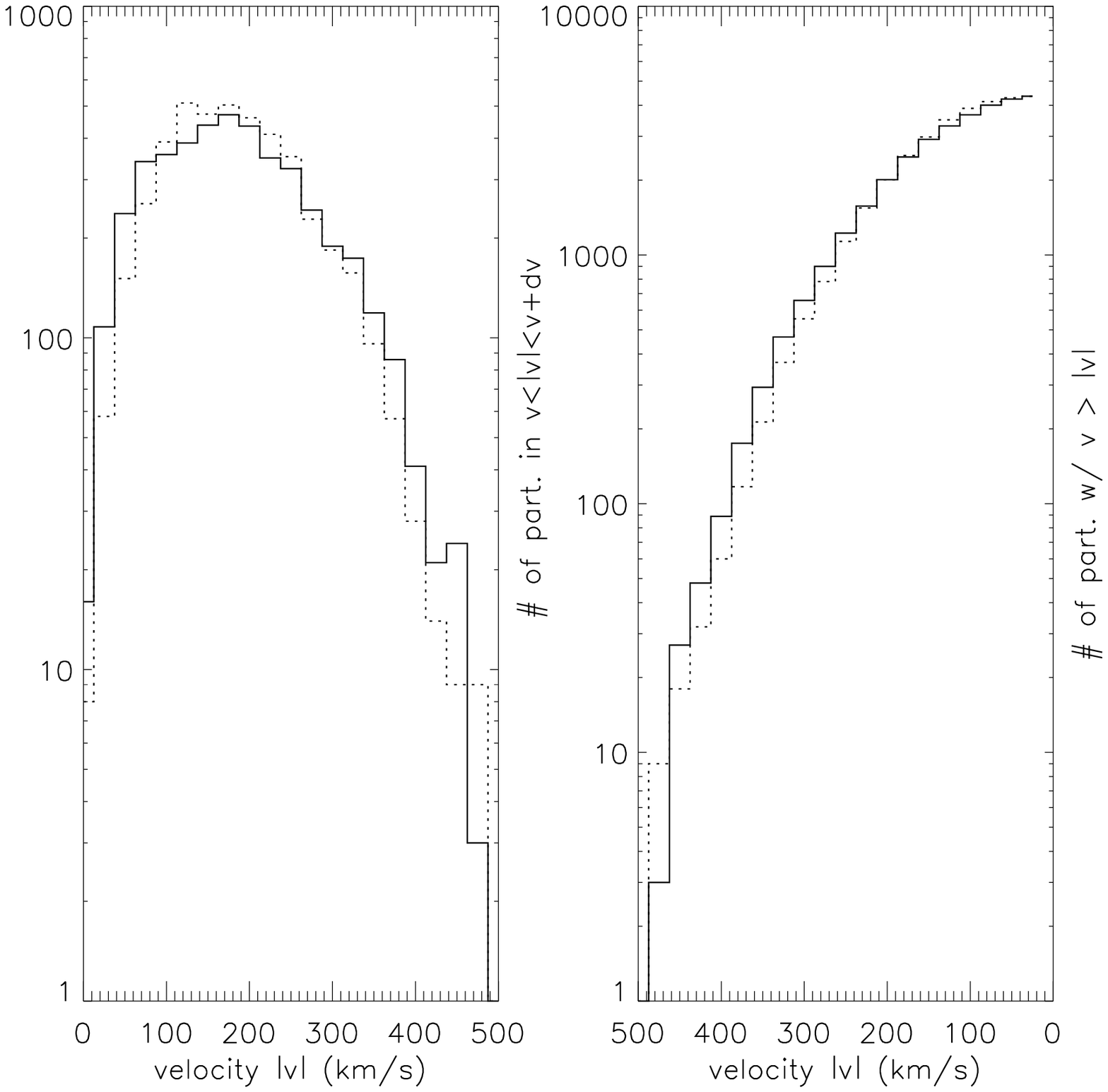,width=7cm}
\psfig{figure=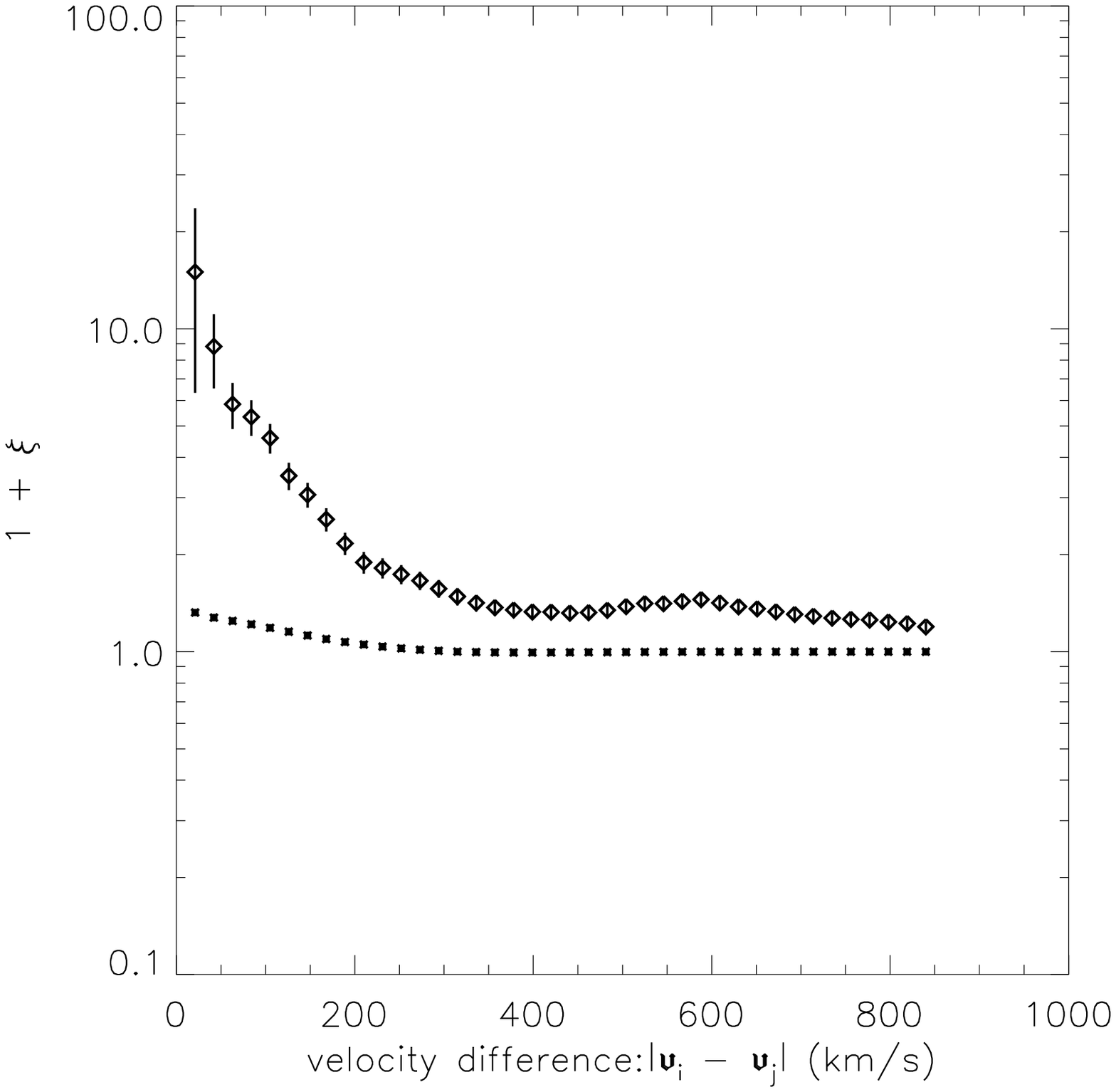,height=6.5cm}}}
\caption{For the same box as in Figure~2 we plot the differential
(left) and the cumulative (centre) velocity distributions.  The dotted
histograms correspond to a multivariate Gaussian. On the right, we
plot the ``correlation function" $\xi$ as the number of neighbours
with velocity differences in a given range compared to what is
expected for random deviates drawn from a multivariate
Gaussian. Asterisks correspond to $\xi$ over all the particles in the
box, whereas diamonds to the 1\% fastest moving particles.}
\end{figure}

\end{document}